\documentclass[12pt]{article}
\usepackage[sc]{mathpazo}
\usepackage[T1]{fontenc}
\usepackage{textcomp}
\usepackage{graphicx}
\usepackage{floatrow}
\usepackage{multicol}
\usepackage[margin=1in]{geometry}

\title{Precise Near-Infrared Radial Velocities with iSHELL}
\author{\small{Bryson Cale (George Mason; bcale@masonlive.gmu.edu), Peter Plavchan (George Mason),}\\\small{Jonathan Gagn\'{e} (Carnegie DTM), Eric Gaidos (U Hawaii),}\\\small{Angelle Tanner (Mississippi State), Peter Gao (UC Berkeley)}}

\date{\vspace{-5ex}}

\begin{document}

\maketitle

\begin{abstract}
    \noindent We present a possible NASA key project using the iSHELL near-infrared high-resolution echelle spectrograph on the NASA Infrared Telescope Facility for precise radial velocity follow-up of candidate transiting exoplanets identified by the NASA TESS mission. We briefly review key motivations and challenges with near-infrared radial velocities. We present the current status of our preliminary radial velocity analysis from the first year on sky with iSHELL.
\end{abstract}
\vspace{-1cm}
\section{\textit{Background}}
\vspace{-0.5cm}
Due to the lower masses and cooler surface temperatures of M dwarfs, the reflex motion from orbiting exoplanets is larger, the transit depths larger, and the Habitable Zone (HZ) orbital periods are shorter.  This M dwarf ``opportunity'' as a short-cut to HZ Earth-mass planets was further strengthened by the discoveries from \textit{Kepler} that terrestrial planets in short orbital periods are more common than they are for FGK dwarfs as was predicted from core accretion theory (Dressing \& Charbonneau 2013, Howard et al. 2012, Laughlin et al. 2004), and solidified as transformative science with the discoveries of GJ 1214 b, Prox Cen b and the TRAPPIST-1 system (Charbonneau et al. 2009, Anglada-Escud\'{e} et al. 2016, Gillon et al. 2017). One setback along the way was the realization that the HZ orbital periods for M dwarfs coincided with their 1-10 Gyr rotation periods (Robertson et al. 2014, Vanderburg et al. 2016, Newton et al. 2016).

M dwarfs are the brightest in the near-infrared (NIR), and the flux contrast from cool star-spots is lower than in the visible. The NIR is thus an enticing wavelength regime for M dwarf precise radial velocity (PRV) surveys and for mitigating stellar activity from the rotational modulation of starspots for FGK stars. NIR PRVs, however, historically lag behind visible benchmarks in demonstrated precision (e.g. 5 m/s in Bean et al. 2010 vs. 70 cm/s with HARPS-S/N). Partially due to the lack of instrument stability and precise wavelength calibration, and partially due to the higher costs of spectrographs and detectors in cryogenic vacuum chambers, the science applications of NIR RVs have been relatively limited to studying active and/our young proto-stars at 50--100 m/s precision (e.g. Johns-Krull et al. 2016, Crockett et al. 2012, Bailey et al. 2012, Setiawan et al. 2008). 

The next generation of NIR PRV spectrometers are coming online such as CARMENES and HPF (Wright \& Robertson 2017) with science goals to survey the nearest several hundred M dwarf stars at 1--5 m/s. Preliminary results from CARMENES demonstrate that the RV information content for M dwarfs is somewhat lower than expected from synthetic stellar models in the J- and H-bands (Reiners et al. 2018). However, there remains an island of high RV information content in K-band due to the 2.30 $\mu$m CO bandhead for not just M dwarfs but also late G and K dwarfs (Figure 1).

\begin{figure}[h]
    \floatbox[{\capbeside\thisfloatsetup{capbesideposition={right,center},capbesidewidth=6cm}}]{figure}[\FBwidth]
    {\caption{RV precision in m/s as a function of effective temperature and wavelength for a fixed amount of observing time (60 sec). The hashed regions correspond to wavelengths where the atmosphere is too opaque. While the best result is achieved from 400-600 nm, an opportunity for cooler M dwarfs appears at 2.3 microns due to the CO bandheads.}}
    {\includegraphics[width=8cm]{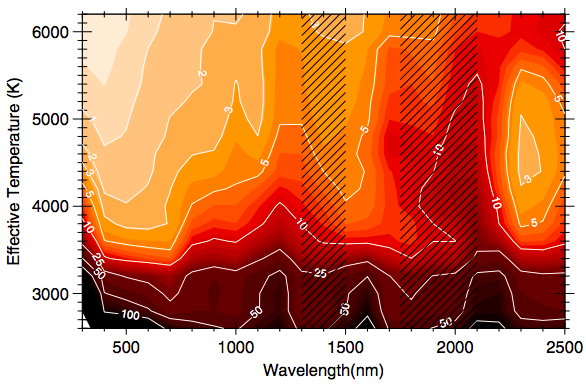}}
\end{figure}

Encouragingly, initial empirical results from CARMENES confirm simulations from toy models and more recently the stellar activity modeling code StarSIM 2.0 that the rotational modulation of cool spots and plages can be detected in the wavelength-dependence (aka \textit{color}) of radial velocity time-series (Figure 2, Tal-Or et al. 2018, Herrero et al. 2016, Reiners et al. 2010).

\begin{figure}[h]
    {\caption{Left: The StarSIM 2.0 simulated RV signals produced from a single, static starspot in the visible and the NIR arm of CARMENES over one rotation period. The amplitude is lower in the NIR, and there is also a phase offset and shape change due to the wavelength dependence of convective blueshift and limb-darkening.  Right: The StarSIM 2.0 simulated RV signal produced from a spotted star as it rotates and the spots evolve in time. The vertical and horizontal axes are the RV signals from visible and NIR arms of CARMENES, respectively. Note that an RV signal from a planet would correspond to a 1:1 line. RV colors also open a new window into stellar astrophysics by breaking the spot size and temperature contrast degeneracy.}}
    {\includegraphics[width=8cm]{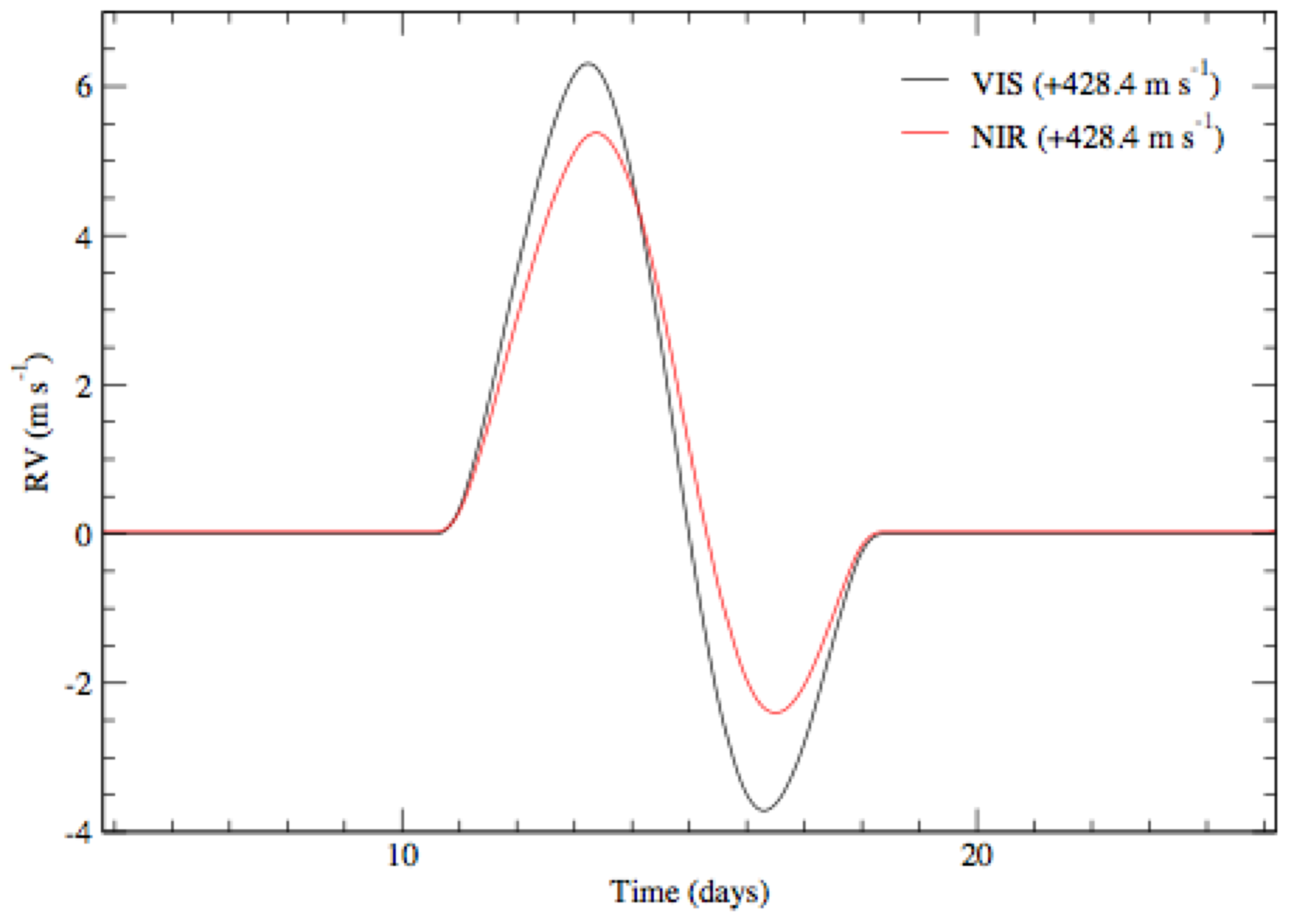}}
    {\includegraphics[width=8cm]{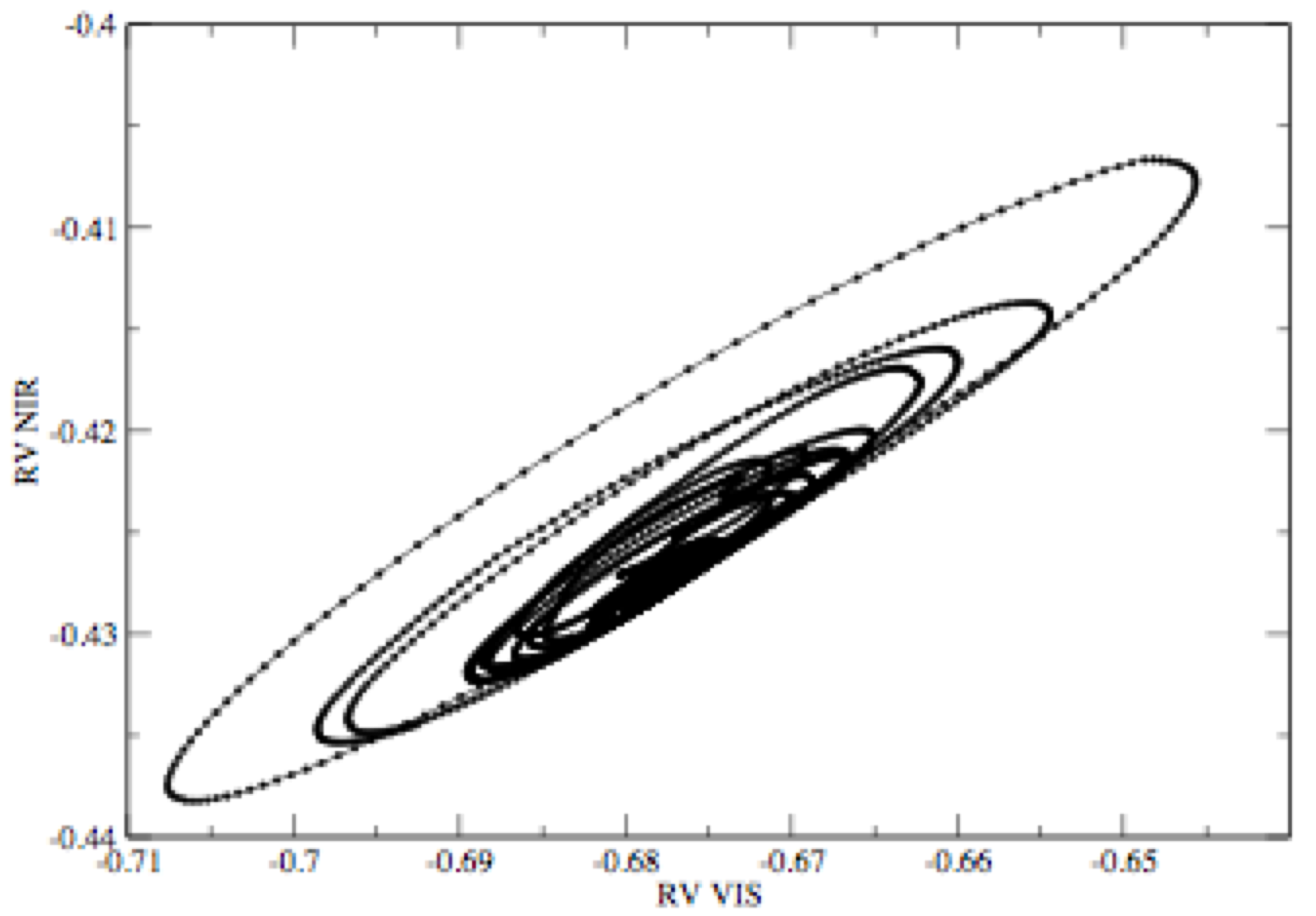}}
\end{figure}

\section{\textit{The iSHELL Spectrograph for NIR PRVs}}
\vspace{-0.5cm}
iSHELL is a 1.1-5.3 $\mu$m high-resolution cross-dispersed echelle spectrograph, replacing the 25-year-old CSHELL spectrograph with improved resolution (R=46k vs. 70k) spectral grasp ( 5nm vs 250 nm), throughput, optics, and detector (Rayner et al. 2012). It is installed on the NASA Infrared Telescope Facility atop Mauna Kea, Hawaii.  First light was achieved in September 2016. We built an isotopic methane gas (13-CH\textsubscript{4}) cell for precise wavelength calibration in the K-band, the same cell that was used in CSHELL (Gagne et al. 2016, Gao et al. 2016, Plavchan et al. 2013, Anglada-Escud\'{e} et al. 2012).  With CSHELL we demonstrated 3 m/s precision over several nights on the red supergiant SV Peg, and 25 m/s long-term precision on nearby M dwarfs, limited primarily by the spectral grasp.

\textbf{IRTF is the only ground-based telescope owned entirely by NASA with a 3-m or larger aperture} with a directive to provide ground-based support for NASA mission objectives, with most of its current funding support from the identification of hazardous near-Earth asteroids (NASA also owns shares of Keck and WIYN telescope observing time and contributes to the interferometer on LBT). \textbf{iSHELL is also the only instrument capable of NIR PRVs with open public access to US PIs.}  Additionally, many of the new red and NIR PRV spectrometers surveying nearby M dwarfs do not go out to K-band, thus creating a unique wavelength range niche for using iSHELL for PRVs calibrated by our gas cell (Figure 3).

\vspace{-0.8cm}
\begin{figure}[h]
    \floatbox[{\capbeside\thisfloatsetup{capbesideposition={left,center},capbesidewidth=6cm}}]{figure}[\FBwidth]
    {\caption{The sensitivity of iSHELL vs. its predecessor CSHELL to HZ exoplanets as a function of stellar mass. A SNR=200 iSHELL spectrum corresponds to a photon noise of approximately 3 m/s, our goal precision. This SNR is reached in 5 hours at K=8.9 mag, and 5 m/s in 1.8 hours.}}
    {\includegraphics[width=8cm]{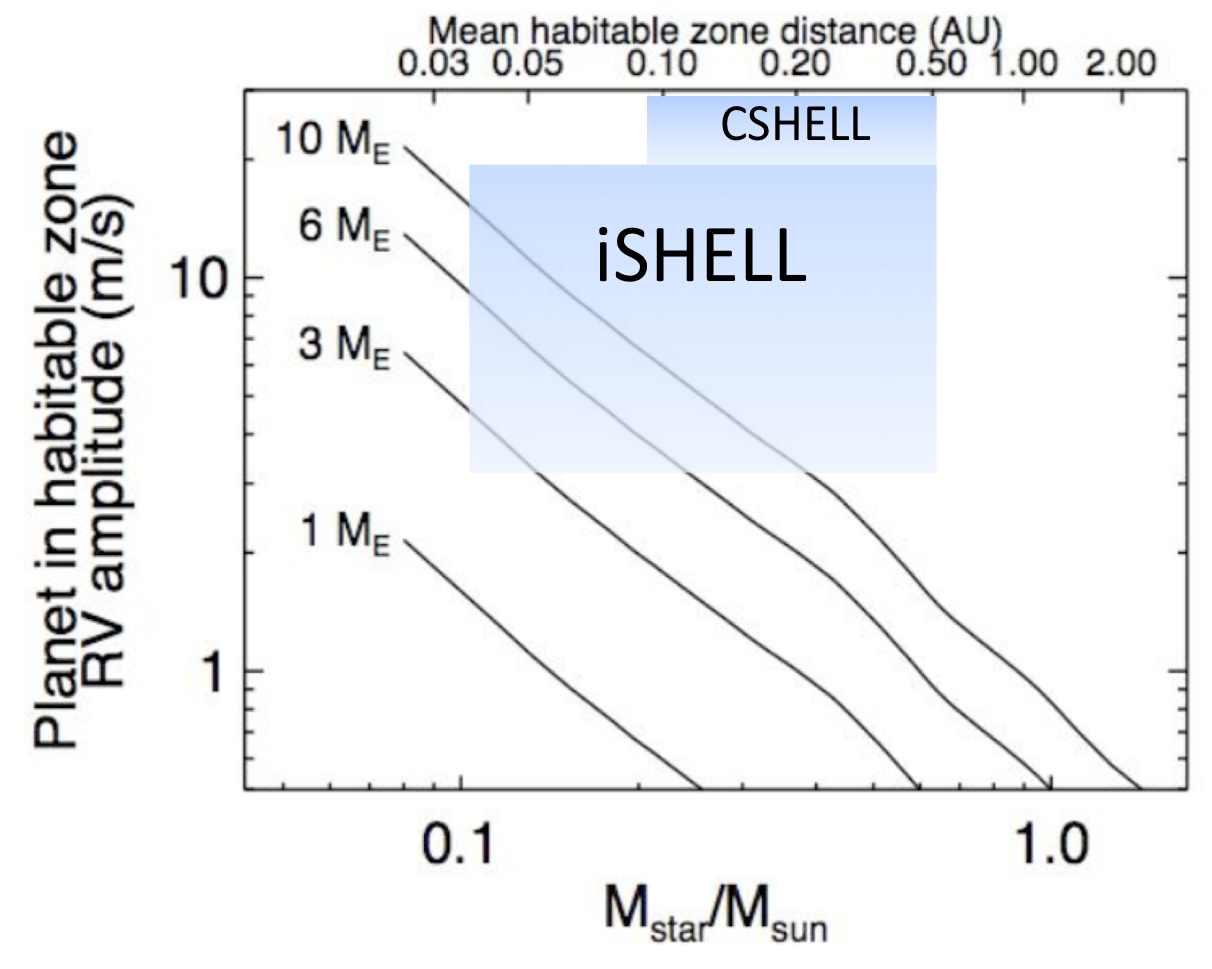}}
\end{figure}
\vspace{-1.8cm}
\section{\textit{Current Status}}
\vspace{-0.5cm}
As of February 2018, we have observed for 25 fractional nights with iSHELL, targeting bright GKM dwarf RV standard stars. Currently, we are focused on demonstrating the achievable RV precision with iSHELL and the methane isotopologue gas cell. We have rewritten our CSHELL data reduction and forward model codes for iSHELL, and we are currently optimizing the codes. The forward model code is based upon the ``grand solution'' first proposed by Jeff Valenti, and implemented in Gao et al. (2016) for CSHELL, where we used the residuals of all of our spectral model fits for a single star, de-shift them to the barycentric rest frame, and co-add them to iteratively improve the stellar template in the presence of the deep NIR telluric (and gas cell) lines.

Our best achieved precision to date with iSHELL is shown in Figure 4 for Barnard's Star (GJ 699), corresponding to a 17 m/s long-term precision using two orders of the 29-order spectrum. On time-scales of less than one week, we have achieved a precision of 3--8 m/s using ten orders for Barnard's Star.  While we have not yet achieved our goal precision of 3 m/s on time-scales of greater than one week, we have identified the likely sources of our systematic velocity errors and are actively working on our modeling code to correct for them. 
\vspace{-0.25cm}
\begin{figure}[h]
    \floatbox[{\capbeside\thisfloatsetup{capbesideposition={right,center},capbesidewidth=6cm}}]{figure}[\FBwidth]
    {\caption{The first RVs obtained from our code for two orders of spectra of Barnard's star (GJ-699). GJ-699 is known to be stable down to 80 cm/s (Anglada-Escud\'{e} \& Butler 2012). Observations were taken on 16 separate nights.  The long-term precision is 17 m/s, with 3--8 m/s obtained for time-scales of less than one week.}}
    {\includegraphics[width=0.5\textwidth]{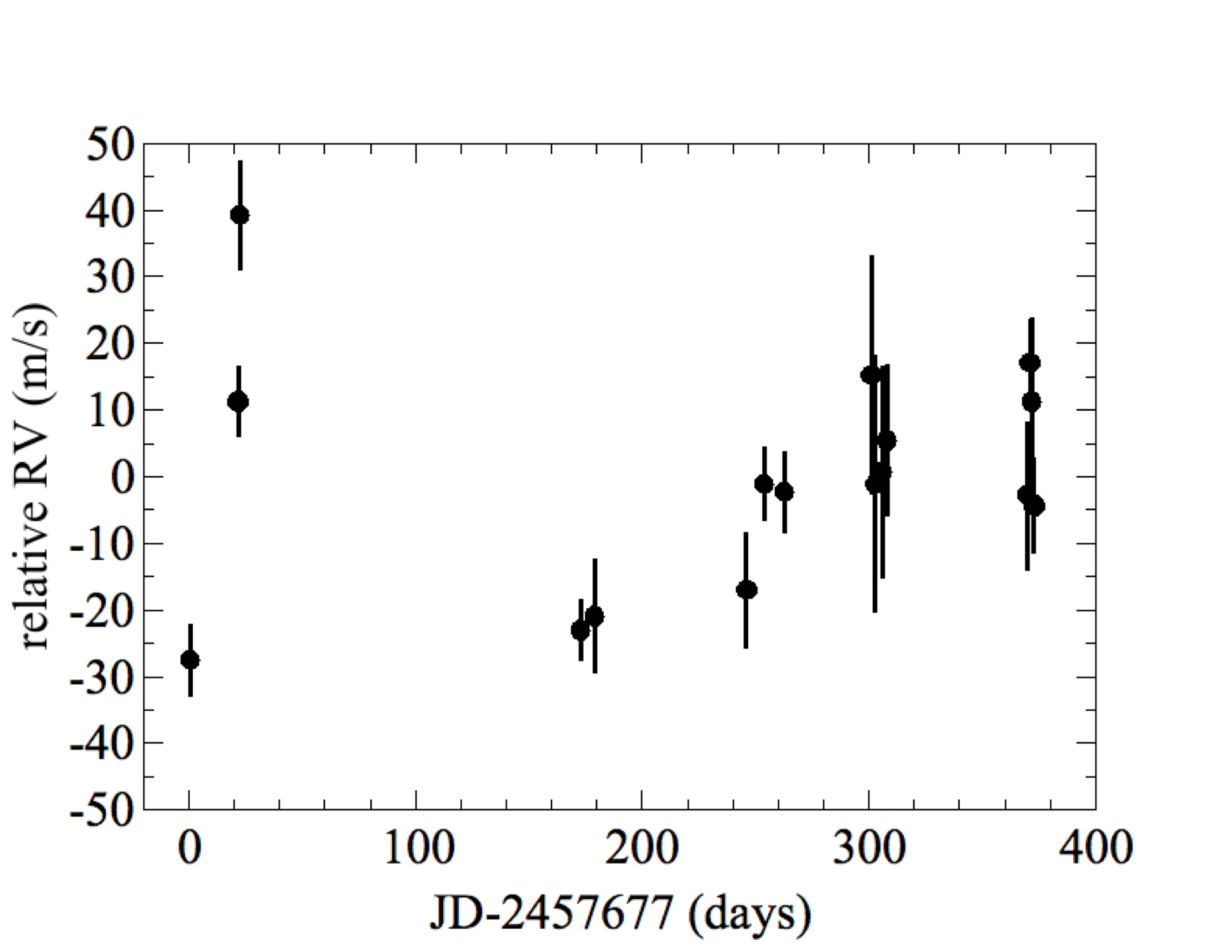}}
\end{figure}

Unfortunately, inherent to the iSHELL data is fringing from several sources. The largest amplitude source of fringing is from a filter in the optical path, and an order is being place to replace the filter.  In the RVs for Barnard's Star in Figure 4, we did not include a fringing component in our forward model. We are currently in the process of testing the performance of the fringing models in our forward model code.
\vspace{-0.5cm}
\section{\textit{TESS Follow-up with iSHELL}}
\vspace{-0.5cm}
The NASA TESS mission is set to launch in the spring of 2018 with the first data release likely to come in early 2019. TESS has the Level 1 mission requirement to measure the masses of 50 planets smaller than 4 R$_\oplus$ coordinated through the TFOP working group and comprising 4000+ hours of dedicated PRV observing time (Ricker et al. 2015). While this may be sufficient to meet the primary TESS goal of determining the masses of 50 planets, it is not utilizing TESS to its full extent where many interesting targets will be left out. Current secured resources only allow for follow-up observations of 1\% of the approximately 5000 candidate exoplanets it will likely identify for RV semi-amplitudes of 1.5-5 m/s (Sullivan et al. 2015).

\textbf{We strongly recommend that NASA implement a key project (e.g. 10-20\% of the time, 100 nights) on a telescope that it already owns with an instrument already on sky and already capable of 20 m/s long-term NIR RVs and 3--8 m/s RVs on time-scales of less than a week to: confirm and measure the masses of up to an additional 50 TESS candidate exoplanets, to identify and/or confirm higher-mass Neptune and Jovian planets, and to perform single-epoch reconnaissance spectroscopy of candidates.} This precision is sufficient to confirm the masses of super-Earths with radii of $<$3 R$_\oplus$ with a brightness of K$<$10 mag and a velocity semi-amplitude of K $>$ 3 m/s (Figure 5, I$<$12, since the typical M dwarf has I-K$\sim$2). For orbital periods of greater than one week, even if we don't achieve our goal of 3 m/s long-term RV precision, iSHELL can be utilized to identify and/or confirm higher mass Neptune and Jovian planet candidates found by TESS.

\begin{figure}[h]
    {\caption{TESS yield of Super-Earths from Sullivan et al. 2015 as a function of apparent magnitude and velocity semi-amplitude. Greyed out regions are either too faint for iSHELL or below our goal precision of 3 m/s.}}
    {\includegraphics[width=1.0\textwidth]{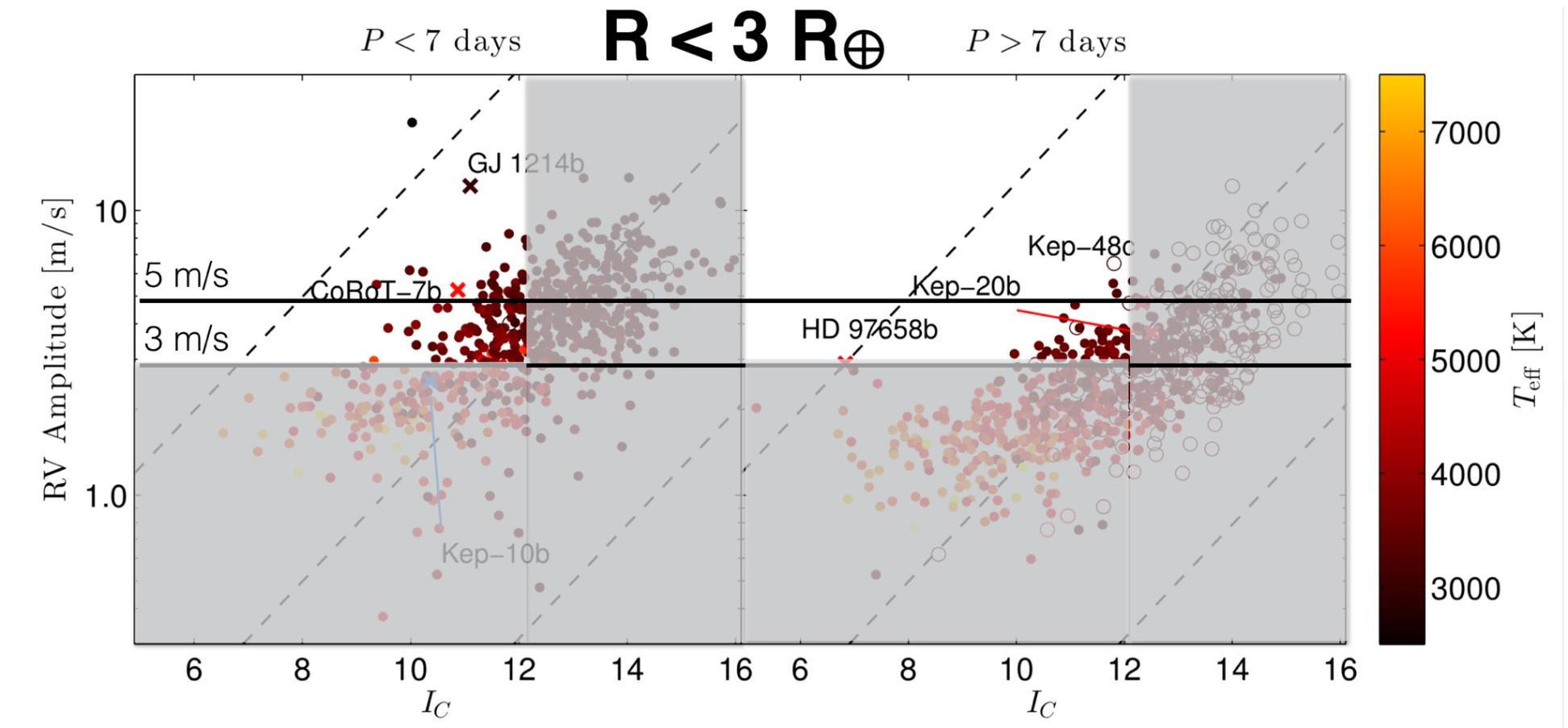}}
\end{figure}

\vspace{-0.5cm}
\section{\textit{References}}
\vspace{-0.75cm}
\begin{multicols}{2}
\small{
\noindent Anglade-Escud\'{e} et al., 2012, PASP, 124, 916 \\
Anglade-Escud\'{e} et al., 2016, Nature, 536, 7617 \\
Anglade-Escud\'{e}, G. \& Butler, P, 2012, ApJS, 200, 15 \\
Bailey et al., 2012, ApJ, 749, 16, 15 \\
Bean et al., 2010, ApJ, 713, 1 \\
Bottom et al., 2013, PASP, 125, 925 \\
Charbonneau et al., 2009, Nature, 462, 7275 \\
Crockett et al., 2012, ApJ, 761, 2 \\
Dressing, C. \& Charbonneau, D., 2013 ApJ, 767, 1 \\
Fischer et al., 2016, 128, 964 \\
Gagn\'{e} et al., 2016, ApJ, 822, 1 \\
Gao et al., 2016, PASP, 128, 968 \\
Gillon et al., 2017, Nature, 542, 7642 \\
Howard et al., 2012, 201, 2, ApJ \\}
\columnbreak
\small{
Johns-Krull et al., 2016, ApJ, 826, 2 \\
Herrero et al., 2015, A\&A, 586, A131 \\
Laughlin, G., Bodenheimer, P., Adams, F., 2004, ApJ, 612, L73 \\
Plavchan et al., 2013, SPIE, 8864, id. 88641J 19 \\
Rayner et al., 2012, SPIE, 8446, id. 84462C 12 \\
Ricker et al., 2015, JAT, Volume 1, id. 014003 \\
Reiners et al., 2010, ApJ, 710, 1 \\
Reiners et al., 2017, A\&A, in press, arXiv:1711.06576 \\
Ricker, RE, Winn, JN, Vanderspek, R., et al., 2015, J. Astron. Telesc. Instrum. Syst., 1, 1 \\
Robertson et al., 2014, Science, 345, 6195 \\
Setiawan et al., 2008, Nature, 451, 7174 \\
Sullivan et al., 2015, 809, 77, 29 \\
Vanderburg et al., 2016, MNRAS, 459, 4 \\
Wright, J. \& Robertson, P., 2017, RNAAS, 1, 1}
\end{multicols}
\end{document}